\documentclass{article}
\usepackage{spconf,amsmath,amssymb,graphicx,hyperref,booktabs,xcolor}
\usepackage{CJKutf8,float}
\newcommand{\papertablestyle}{%
  \small
  \setlength{\tabcolsep}{2pt}%
  \renewcommand{\arraystretch}{1.12}%
  \setlength{\heavyrulewidth}{0.6pt}%
  \setlength{\lightrulewidth}{0.35pt}%
  \setlength{\cmidrulewidth}{0.35pt}%
}
\newcommand{\tablenote}[1]{\par\vspace{3pt}%
  {\small\raggedright #1\par}}

\title{A LEARNED MUSICAL-TIME COORDINATE FOR PIANO AUDIO-TO-SCORE TRANSCRIPTION}
\name{MIN-HSING WANG, TSUNG-REN HUANG}
\address{National Taiwan University\\\texttt{\{mhwang,tren\}@mil.psy.ntu.edu.tw}}

\begin{document}
\ninept
\maketitle

\begin{abstract}
End-to-end piano audio-to-score (A2S) transcription is the task of reconstructing a notated score from an audio recording in a single system. Existing A2S approaches usually rely on pre-segmented excerpts, and few handle whole pieces without external segmentation. This raises the question of whether a musical-time coordinate can guide score transcription while also helping the system segment whole pieces. Previous work on jazz saxophone A2S uses an external beat tracker to obtain such a coordinate for score prediction. We introduce downbeat phase, a musical-time coordinate learned jointly with score transcription that represents each audio frame's relative position within a notated bar. With this coordinate, our system can not only condition its score decoder on downbeat phase but also self-segment at bar boundaries and stitch the segments into a whole piece. Evaluation suggests that the coordinate improves metrical structure and supports whole-piece transcription with comparable notation quality.
\end{abstract}


\section{Introduction}
\label{sec:intro}

Audio-to-score (A2S) transcription aims to learn how to recover a text-based score from a music audio recording. A score represents pitches and organizes their timing through note values, beats, and bars. A performer not only plays the notes but also determines the tempo throughout the piece and shapes how these notated rhythms unfold in time. Therefore, transcribing scores from performances involves inferring the notated pitch and duration of notes and their timing within each bar, and predicting bar-level information such as time and key signatures. Among existing end-to-end A2S systems, Piano-A2S, proposed by Zeng et al.~\cite{zeng2024endtoend}, directly transcribes piano audio recordings through a GRU decoder, predicting bar-level and note-level information in a hierarchical manner. Alfaro-Contreras et al.~\cite{alfaro-contreras2024transformer} use a Transformer decoder to derive scores from polyphonic string music. However, prior work mainly focuses on score sequence prediction from excerpts spanning several bars and pre-segmented at bar boundaries. How to assemble the transcribed score segments into a whole piece remains mostly unexplored.

To extend transcription to whole pieces, it is worth exploring whether a musical-time coordinate can support an A2S system in both score prediction and self-segmentation. Building on the phase-based approach of Oyama et al.~\cite{oyama2021phaseaware}, Shanin et al.~\cite{shanin2026audiotoscore} express position within each externally estimated beat as a circular phase and encode it with Fourier features to guide jazz saxophone score transcription. This leaves open whether the coordinate itself can be learned jointly with score transcription.

\begin{figure*}[t]
\centering
\includegraphics[width=0.9\textwidth]{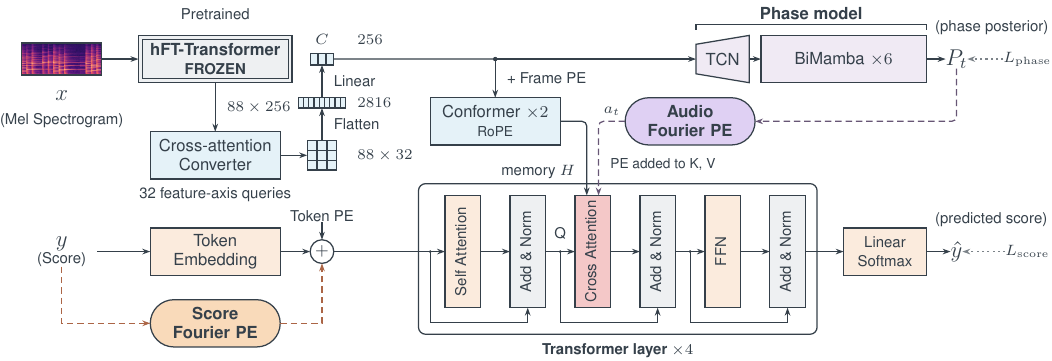}
\par\vspace{-6pt}
\caption{The encoder extracts content features and estimates downbeat phase, while the decoder combines these features with audio- and score-side Fourier positional encodings of downbeat phase to generate a score autoregressively.}
\label{fig:system}
\end{figure*}

We introduce an end-to-end piano A2S system that learns a musical-time coordinate jointly with score transcription.\footnote{To reproduce our settings, the implementation is available at \url{https://github.com/bloggerwang1217/MusicTimeA2S}.} As shown in Fig.~\ref{fig:system}, this coordinate is represented as downbeat phase and supplied to the score decoder through Fourier positional encoding. The predicted phase posterior also provides bar boundaries, allowing the system to self-segment and stitch the segments into a whole piece. We evaluate the system under pre-segmented and whole-piece transcription protocols. Pre-segmented evaluation shows that the learned coordinate helps capture metrical structure. In whole-piece transcription, our system achieves higher metrical scores and comparable or better notation quality without external segmentation.

\section{Method}
\label{sec:method}

Given a piano recording represented as a log-mel spectrogram $x \in \mathbb{R}^{T \times M}$, our system generates a score token sequence $\hat{y}$. Following the preprocessing settings of hFT-Transformer~\cite{toyama2023automatic}, we use audio sampled at 16\,kHz with $M=256$ mel bins and a temporal resolution of 16\,ms per frame. Each input window contains up to $T=1{,}280$ frames, covering 20.48\,s of audio.

The target score $y=(y_1,\ldots,y_N)$ is a sequence of $N$ tokens drawn from a vocabulary $\mathcal{V}$ of 248 tokens. Its representation is derived from the \texttt{**kern} format used by Piano-A2S~\cite{zeng2024endtoend} and encodes pitches, note values, time and key signatures, and structural markers for bars and voices. When tokenizing, we change the serialization order for \texttt{**kern}. Specifically, the token sequence emits the pitch at the note's notated onset and its duration at the notated offset. The decoder predicts these tokens autoregressively to form $\hat{y}$, with a limit of 2,048 tokens per decoding pass.

\subsection{Architecture}

\subsubsection{Encoder}
We use a frozen hFT-Transformer~\cite{toyama2023automatic} pretrained on MAESTRO~\cite{hawthorne2019enabling} as a front-end audio feature extractor. It produces a 256-dimensional feature vector for each of the 88 piano pitches at every audio frame. To obtain compact frame features, a cross-attention converter uses 32 learned queries to summarize the feature-channel axis, producing an $88\times32$ representation. We then flatten and linearly project each frame's representation to a 256-dimensional vector for subsequent temporal modeling.

The resulting sequence $C$ feeds two parallel branches for audio content and downbeat phase. In the content branch, frame positional encoding (Frame PE) assigns each frame a fixed sinusoidal vector determined by its index within the input window. We add these vectors to $C$ before two Conformer blocks~\cite{gulati2020conformer} that produce content memory $H$, with 256 dimensions per frame. Their self-attention uses rotary position embedding (RoPE)~\cite{su_roformerenhancedtransformer_2024} to encode frame positions in the queries and keys.

The downbeat phase branch, which we refer to as the phase model, applies a dilated temporal convolutional network (TCN) followed by six bidirectional Mamba (BiMamba) layers. The TCN follows the seven-block residual design of Oyama et al.~\cite{oyama2021phaseaware}. We adapt BeatMamba's~\cite{ru2026beatmamba} BiMamba layers with Mamba-2~\cite{dao2024transformers} as a fixed-resolution stack. In place of beat-event prediction, a framewise phase classifier estimates downbeat phase. Its output is used to construct the Audio Fourier PE defined in Sec.~\ref{sec:fourierpe}.

\subsubsection{Fourier Positional Encoding of Downbeat Phase}
\label{sec:fourierpe}
To support the prediction of bar boundaries, we draw on the phase-based downbeat estimation of Oyama et al.~\cite{oyama2021phaseaware} and represent downbeat phase by an angle $\phi\in[0,2\pi)$ that completes one cycle per notated bar and resets at each downbeat.

We define $\gamma(\phi)$ as a shared 24-dimensional Fourier feature map for audio- and score-side downbeat phase:
\begin{equation}
	\gamma(\phi)=
	[\sin(h\phi)]_{h=1}^{12}\,\Vert\,
	[\cos(h\phi)]_{h=1}^{12}\in\mathbb{R}^{24},
\end{equation}
where $h$ is the harmonic order and $\Vert$ denotes concatenation. The $h$-th harmonic repeats $h$ times per bar. We retain harmonics up to order 12 to cover the rhythmic units specified by our supported time signatures---for example, twelve eighth-note units per bar in 12/8 time.

At each audio frame $t$, in both training and inference, softmax converts the phase classifier's logits into a posterior $P_t$ over 360 within-bar positions. Here, $P_t(r)$ is the probability assigned to position class $r\in\{0,\ldots,359\}$. We denote the 24-dimensional audio-side Fourier positional encoding of downbeat phase (Audio Fourier PE) at frame $t$ by $a_t$. With class centers $\theta_r=2\pi r/360$, we compute this vector as
\begin{equation}
	a_t=\sum_{r=0}^{359}P_t(r)\gamma(\theta_r).
\end{equation}

\subsubsection{Transformer Decoder}
The decoder is a four-layer Transformer with 256-dimensional hidden states and four attention heads. At decoder input position $j$, the score prefix consumed through the current token determines the current score time within the bar. We define $\phi_j^{\mathrm{score}}$ as $2\pi$ times this time divided by the bar duration specified by the current time signature, wrapped to $[0,2\pi)$. It resets at bar tokens and is set to zero before a time signature is available. Applying the same feature map $\gamma$ gives the 24-dimensional score-side Fourier positional encoding of downbeat phase (Score Fourier PE), $s_j=\gamma(\phi_j^{\mathrm{score}})$, for predicting the next token. The decoder input sums token embeddings, sinusoidal token positional encoding (Token PE) for token order, and a learned projection of $s_j$. In each decoder layer $\ell$, cross-attention keys and values combine audio content with its downbeat phase:
\begin{equation}
\begin{aligned}
k_t^{(\ell)} &= W_K^{(\ell)}H_t+b_K^{(\ell)}+U_K^{(\ell)}a_t,\\
v_t^{(\ell)} &= W_V^{(\ell)}H_t+b_V^{(\ell)}+U_V^{(\ell)}a_t.
\end{aligned}
\end{equation}
Here, $H_t$ is the content memory at audio frame $t$. The $W$ matrices and bias terms project this content into keys and values, while $U_K^{(\ell)},U_V^{(\ell)}\in\mathbb{R}^{256\times24}$ separately project Audio Fourier PE $a_t$ into the same spaces for each layer.
\subsection{Training Objective}
The score sequence is supervised by a teacher-forced next-token cross-entropy loss $L_{\mathrm{score}}$. To learn downbeat phase, we introduce a framewise phase loss following Oyama et al.~\cite{oyama2021phaseaware},
\begin{equation}
L_{\mathrm{phase}}(t)=-\sum_{r=0}^{359}q_t(r)\log P_t(r),
\end{equation}
where $q_t$ is the target distribution and $P_t$ the predicted posterior at frame $t$. The target phase $\phi_t^\ast$ is quantized to the nearest of the 360 circular position classes, then smoothed with triangular circular weights over $\pm3$ classes. Following Shanin et al.~\cite{shanin2026audiotoscore}, we anchor the $k$-th beat of an $m$-beat bar at $2\pi k/m$ and linearly interpolate within each annotated beat to construct $\phi_t^\ast$. The total loss is $L = L_{\mathrm{score}} + \lambda\,L_{\mathrm{phase}}$ with $\lambda = 0.1$, where $L_{\mathrm{phase}}$ is $L_{\mathrm{phase}}(t)$ averaged over all frames with phase annotations.

\subsection{Inference}
\label{sec:inference}
To transcribe the whole recording, we first estimate its
bar boundaries and token sequences of time and key signatures, then generate
and assemble the score. We scan the audio with 1,280-frame windows at a stride
of 640 frames and combine overlapping phase predictions
by averaging their circular first moments.
The downbeats are linearly interpolated forward phase-zero crossings of the averaged first moment,
with the first crossing retained and a minimum interval of 400\,ms
enforced between successive downbeats to reduce over-segmentation.
For each five-bar excerpt, the decoder scores candidate time-signature prefixes
and the key-signature tokens that follow them.
We apply separate first-order Markov models to the token sequences
of time and key signatures across five-bar excerpts.
Viterbi decoding maximizes the sum of decoder log probabilities,
with a penalty for each change.
Key-signature probabilities are marginalized over time signatures.
The time- and key-signature penalties, 8.3 and 6.3, are derived from their
change rates between consecutive five-bar excerpts in the training split.

With both signature sequences fixed, we transcribe each five-bar excerpt
through one or more fixed-length audio windows.
Each window recomputes its local Audio Fourier PE.
Downbeats within each window's first half determine how many bars to decode. After each grammar-constrained greedy attempt, we retain the
longest prefix of complete bars that can be reconstructed
into \texttt{**kern}, then move the window to the following
downbeat and restart decoding with a fresh autoregressive
state. If no bar is accepted, we retry from the preceding downbeat when available
and discard the first bar of the retry. This continues until each
five-bar excerpt is complete. Finally, the excerpts are assembled
into a whole piece, with missing segments represented
by blank bars.

\section{Experiments}
\label{sec:exp}

\subsection{Data}
\label{sec:data}

\noindent\textbf{Datasets.} Three corpora are adopted: HumSyn, MuseSyn,
and ASAP~\cite{foscarin2020asap}.
HumSyn contains classical and ragtime piano \texttt{**kern} scores from the Humdrum
database.\footnote{\url{https://kern.ccarh.org/}}
MuseSyn contains piano arrangements collected from MuseScore in MusicXML
format, mainly of popular music.\footnote{\url{https://musescore.com/hub/piano}}
ASAP provides piano scores aligned with human
performances, including audio with downbeat and beat annotations.

We adopt HumSyn and MuseSyn, jointly referred to as Syn, for training and
validation and use Syn and ASAP for evaluation. The Syn split is based on
that of Zeng et al.~\cite{zeng2024endtoend}, with works used in the ASAP
evaluation excluded. Additionally, for the test split, we exclude works
overlapping with the training or validation data of any evaluated system.
The resulting splits contain 546 scores for training,
46 scores for validation, and 45 scores for testing. For the ASAP dataset, we use 74 recordings from 39 works, taken from
the test split of 102 recordings released by Liu et al.~\cite{liu2022performance}
after excluding recordings used to train our hFT-Transformer front
end~\cite{toyama2023automatic}.

\begin{table}[t]
\centering
\caption{Pre-segmented MV2H evaluation on Syn and ASAP.}
\label{tab:preseg}
\par\vspace{4pt}
\papertablestyle
\setlength{\tabcolsep}{2pt}
\begin{tabular*}{\columnwidth}{@{\extracolsep{\fill}}l@{\hspace{3pt}}rrrrr@{\hspace{5pt}}r@{\hspace{5pt}}r@{\hspace{2pt}}}
\toprule
& \multicolumn{5}{c}{MV2H components} & \multicolumn{2}{c}{MV2H} \\
\cmidrule(lr){2-6}\cmidrule(l){7-8}
System & $F_{\mathrm{mp}}$ & $F_{\mathrm{voi}}$ & $F_{\mathrm{meter}}$ & $F_{\mathrm{val}}$ & $F_{\mathrm{harm}}$ & 4 & 5 \\
\midrule
\multicolumn{8}{@{}l}{\textbf{Syn} (9,958 excerpts, 28 works)} \\
\addlinespace[2pt]
Piano-A2S & .681 & .876 & .352 & .901 & .638 & .774 & .690 \\
\addlinespace[4pt]
Ours & .792\rlap{$^{\ast}$} & .885 & \textbf{.675}\rlap{$^{\dagger}$} & .930\rlap{$^{\ast}$} & .749\rlap{$^{\ast}$} & .839\rlap{$^{\ast}$} & \textbf{.806}\rlap{$^{\ast}$} \\
\hspace{0.6em}w/o coordinate & \textbf{.793}\rlap{$^{\ast}$} & \textbf{.890}\rlap{$^{\ast}$} & .557\rlap{$^{\ast}$} & \textbf{.936}\rlap{$^{\ast}$} & \textbf{.768}\rlap{$^{\ast}$} & \textbf{.847}\rlap{$^{\ast}$} & .789\rlap{$^{\ast}$} \\
\midrule
\multicolumn{8}{@{}l}{\textbf{ASAP} (633 excerpts, 8 works)} \\
\addlinespace[2pt]
Piano-A2S & .636 & \textbf{.882} & .351 & \textbf{.903} & .512 & .733 & .657 \\
\addlinespace[4pt]
Ours & .621 & .833 & \textbf{.384} & .889 & \textbf{.657}\rlap{$^{\ast}$} & .750 & .677 \\
\hspace{0.6em}w/o coordinate & \textbf{.652} & .851 & .357 & .901 & .630\rlap{$^{\ast}$} & \textbf{.759} & \textbf{.678} \\
\bottomrule
\end{tabular*}
\tablenote{Ours rows: mean over three training seeds.\\
$\ast$: significant improvement over Piano-A2S;\\
$\dagger$: significant improvement over all other systems.}
\end{table}

\vspace{3pt}
\begingroup
\emergencystretch=3em
\noindent\textbf{Data augmentation.} We augment the training data by varying
tempo and performance style with VirtuosoNet~\cite{jeong2019virtuosonet}
during MIDI generation, and SoundFonts and reverb during audio rendering.
We scale the reference tempo by a multiplier
$\alpha \in [1/\sqrt{2},\,\sqrt{2}]$, sampled on a logarithmic scale.
Each training score is rendered into four audio versions using different
combinations of four piano SoundFonts and four room impulse responses.
At each training epoch, one version is randomly selected for each score.

The Syn test split follows the synthesis protocol of
Zeng et al.~\cite{zeng2024endtoend}. Each test score has four performance
versions, rendered with a training-held-out SoundFont.
\par\endgroup

\begin{table*}[t]
\centering
\caption{Whole-piece transcription quality on Syn and ASAP.}
\label{tab:selfseg}
\par\vspace{1pt}
\papertablestyle
\newcommand{\refcell}[1]{\textcolor{black!65}{#1}}
\begin{tabular*}{\textwidth}{@{\extracolsep{\fill}}lllrrrrrrrr@{}}
\toprule
& & & \multicolumn{4}{c}{\textbf{Syn} (180 renderings, 45 works)} & \multicolumn{4}{c}{\textbf{ASAP} (74 recordings, 39 works)} \\
\cmidrule(lr){4-7}\cmidrule(lr){8-11}
System & \begin{tabular}[b]{@{}l@{}}External\\Segmentation\end{tabular} & Coordinate & $F_{\mathrm{meter}}$ & $\mathrm{MV2H}_5$ & OMR-NED & Transcribed (\%)
       & $F_{\mathrm{meter}}$ & $\mathrm{MV2H}_5$ & OMR-NED & Transcribed (\%) \\
\midrule
Piano-A2S & Beat This! & n/a
& .236 & .614 & 75.4 & 98
& .143 & .584 & 84.0 & 97 \\
& \refcell{oracle} & \refcell{n/a}
& \refcell{.337} & \refcell{.650} & \refcell{67.9} & \refcell{77}
& \refcell{.168} & \refcell{.566} & \refcell{72.6} & \refcell{78} \\
\addlinespace[2pt]
Ours & --- & predicted
& .667\rlap{$^{\ast}$} & .829\rlap{$^{\ast}$} & 61.8\rlap{$^{\ast}$} & 100
& .318\rlap{$^{\ast}$} & .718\rlap{$^{\ast}$} & 85.4 & 100 \\
& \refcell{oracle} & \refcell{predicted}
& \refcell{.696\rlap{$^{\ast}$}} & \refcell{.837\rlap{$^{\ast}$}} & \refcell{60.0\rlap{$^{\ast}$}} & \refcell{100}
& \refcell{.287\rlap{$^{\ast}$}} & \refcell{.701\rlap{$^{\ast}$}} & \refcell{85.1} & \refcell{100} \\
& \refcell{oracle} & \refcell{oracle}
& \refcell{.807\rlap{$^{\dagger}$}} & \refcell{.870\rlap{$^{\ast}$}} & \refcell{53.8\rlap{$^{\ast}$}} & \refcell{100}
& \refcell{.518\rlap{$^{\dagger}$}} & \refcell{.772\rlap{$^{\dagger}$}} & \refcell{75.1\rlap{$^{\ast}$}} & \refcell{100} \\
\bottomrule
\end{tabular*}
\tablenote{Ours rows: mean over three training seeds. $\ast$: significant improvement of Ours over Piano-A2S with Beat This! downbeat segmentation; $\dagger$: significant improvement of Ours over all other systems. MV2H and OMR-NED are computed on each system's successfully transcribed recordings, with missing segments in the output scores filled with blank bars.}
\end{table*}
\subsection{Training Configuration}

The full model has 23.4M parameters: 17.9M trainable and 5.5M frozen in the hFT-Transformer.
Our models are trained for a maximum of 50 epochs.
The best checkpoints are selected
by validation $L_{\mathrm{score}}$ with an early stopping patience
of five epochs. We use an AdamW optimizer with weight decay of 0.01 and dropout
at a rate of 0.1.
The learning rate warms up linearly to $10^{-4}$ over 1,000 steps and
follows cosine annealing to $10^{-5}$ at epoch 50; the phase model's learning
rate instead anneals from step 10,000 to 20,000 and stays fixed thereafter.

\subsection{Metrics}
\label{sec:metrics}

We evaluate transcription quality using MV2H and OMR-NED. Significance is assessed with 10,000 paired bootstrap draws over works, independently resampling training seeds for each system. A difference is considered significant when its 95\% bootstrap confidence interval excludes zero.

\vspace{3pt}
\begingroup
\noindent\textbf{MV2H}~\cite{mcleod2018evaluating}.
This metric evaluates score transcription
across five aspects: multi-pitch detection ($F_{\mathrm{mp}}$), voice separation
($F_{\mathrm{voi}}$), metrical position ($F_{\mathrm{meter}}$),
note values ($F_{\mathrm{val}}$), and harmony ($F_{\mathrm{harm}}$).
$F_{\mathrm{meter}}$ is most relevant to our task. It is the F1 score of bar,
beat, and sub-beat groupings that start and end at the same notes as in the
reference, measuring where notes sit within bars and beats.
$\mathrm{MV2H}_5$ is the mean of all five components.
Following Zeng et al.~\cite{zeng2024endtoend}, Table~\ref{tab:preseg} also reports
$\mathrm{MV2H}_4$, which averages all components except $F_{\mathrm{meter}}$.

Ground-truth score MIDI is required as a reference for MV2H evaluation. The pre-segmented experiment uses five-bar excerpts and reference MIDI from Zeng et al.'s pipeline~\cite{zeng2024endtoend}. Coordinate ablation and whole-piece MV2H evaluation use reference MIDI from our preprocessing pipeline.

We use the official implementation of non-aligned
MV2H~\cite{mcleod2019evaluating} for pre-segmented transcription.
Because its alignment is prohibitively slow for whole
pieces~\cite{beyer2024endtoenda}, we implement a single-path version that
breaks ties by chord matches~\cite{cogliati2017metric} for self-segmented
transcription, and assess its consistency with the official implementation on
pre-segmented excerpts.\footnote{Mean absolute per-excerpt $\mathrm{MV2H}_5$ difference: 0.0059 (Syn), 0.0076 (ASAP); largest in $F_{\mathrm{meter}}$: 0.0215, 0.0275.}

\vspace{3pt}
\noindent\textbf{OMR-NED}~\cite{martinez-sevilla2025sheet}.
This metric counts music-symbol insertions and deletions between predicted
and reference scores; lower is better. We use it for whole-piece
reconstruction quality.
\par\endgroup

\subsection{Results from Pre-segmented Transcription}
\label{sec:preseg}

For Piano-A2S, we use the ASAP-fine-tuned version from Zeng et al.~\cite{zeng2024endtoend}.
For the ASAP test split, we exclude works appearing in Piano-A2S training or validation data.
Within each dataset, we use the common excerpts scored for all evaluated models.

In Table~\ref{tab:preseg}, both of our model variants
score higher than Piano-A2S on every component on Syn. Compared with the variant
without the coordinate (Ours w/o coordinate), the full model achieves its largest gain in
$F_{\mathrm{meter}}$, with a significant improvement. The other component
scores are slightly lower, although these differences are not significant.
On ASAP, $F_{\mathrm{meter}}$ is slightly higher with the coordinate while
multi-pitch, voice, and note-value scores are lower, leaving $\mathrm{MV2H}_5$
almost unchanged. These results support the coordinate's contribution to
recovering metrical structure on Syn, while its benefit for pre-segmented
transcription on ASAP remains inconclusive. Although introducing the coordinate
slightly lowers some component scores, it also enables self-segmentation for
whole-piece transcription, a trade-off we examine in Sec.~\ref{sec:selfseg}.

\begin{figure}[H]
\centering
\includegraphics[width=\columnwidth]{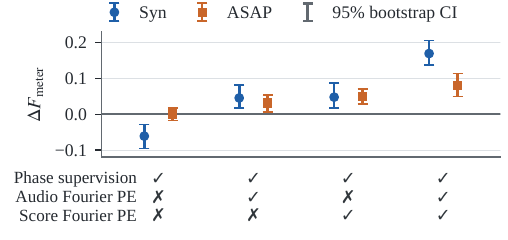}
\par\vspace{-9pt}
\caption{Coordinate ablation using one training seed on common scored excerpts from the test splits in Sec.~\ref{sec:data}. $\Delta F_{\mathrm{meter}}$ denotes the difference from Ours w/o coordinate.}
\label{fig:fourierpe-ablation}
\end{figure}

Fig.~\ref{fig:fourierpe-ablation} presents a coordinate ablation of our full model.
The results show that phase supervision alone does not improve $F_{\mathrm{meter}}$,
and even significantly decreases it on Syn.
While Audio or Score Fourier PE alone significantly improves $F_{\mathrm{meter}}$
on both test sets, the full model combining both achieves the largest gain.

\makeatletter
\@startsection{subsection}{2}{\z@}{-15pt}
  {1.5ex \@plus .2ex}{\normalfont\large\bfseries}
  {Results from Self-segmented Whole-piece Transcription}
\makeatother
\label{sec:selfseg}

Beyond conditioning score decoding, the learned coordinate also provides bar boundaries for whole-piece transcription.
Piano-A2S requires external bar boundaries, which we obtain from Beat This!~\cite{foscarin2024beata}, a state-of-the-art beat tracker evaluated on ASAP.

In Table~\ref{tab:selfseg}, our system achieves significantly higher
$F_{\mathrm{meter}}$ and $\mathrm{MV2H}_5$ than Piano-A2S with Beat This!
segmentation on both test sets. OMR-NED is significantly lower
on Syn and comparable on ASAP. Our system does so without
external segmentation and transcribes all recordings.
The advantage in $F_{\mathrm{meter}}$ and $\mathrm{MV2H}_5$ remains on both
test sets when Piano-A2S uses annotated downbeats.\footnote{Piano-A2S accepts 4--12\,s inputs, so longer annotated segments reduce transcription coverage.}

To examine boundary and phase estimation errors in our system, we evaluate
the oracle conditions shown in gray. In the oracle-coordinate condition,
only Audio Fourier PE uses annotated phase targets.
The results show that replacing the boundaries alone does not consistently improve
metrical scores. Using annotated coordinates as well as boundaries gives a
larger improvement, suggesting room for better coordinate estimation on human
performances. Although the frozen front end is pretrained on real piano
recordings, the phase and score models are trained on synthesized performances.
Generalization may therefore depend on learning expressive timing and its
relation to the notated metrical structure.

\makeatletter
\@startsection{section}{1}{\z@}{-14pt}
  {1.2ex \@plus .2ex}{\normalfont\large\bfseries\centering}
  {\MakeUppercase{Conclusion}}
\makeatother
\label{sec:conclusion}
We present a jointly trained A2S system that uses learned downbeat phase to
condition score decoding and provide boundaries for whole-piece transcription.
Our experiments suggest that supplying this coordinate to the decoder helps
capture metrical structure. Moreover, in self-segmented transcription, bar
boundaries derived from the same coordinate let the system stitch bar
segments into a whole piece without an external beat tracker. Future work will
investigate how expressive timing and notational conventions affect coordinate
estimation, and explore editorial style conditioning to improve the transcription
of expressive piano performances.

\clearpage
\bibliographystyle{IEEEbib}
\bibliography{ICASSP}
\end{document}